\documentstyle[12pt]{article}
\author{S.\ M.\ Troshin and
 N.\ E.\ Tyurin\\
Institute for High Energy Physics\\
Protvino, Moscow Region, 142284 Russia}
\title{Strangeness in  constituent quarks and
one-spin asymmetry in inclusive $\varphi$-meson production}
 \rm \date{}
\begin{document} \maketitle \begin{abstract}
On the basis of the mechanism proposed earlier for one-spin
asymmetries in inclusive hadron production we consider OZI--suppressed
 process of $\varphi$-meson production $pp\rightarrow \varphi X$ and
asymmetry $A_N(\varphi)$ in this process. The main role in
generation of this asymmetry belongs to the orbital angular momentum
of strange quark-antiquark cloud in  internal structure of
constituent quarks. \end{abstract}

A significant experimental information
on spin structure of proton
had been
obtained by the present time.
As it is
considered now, about one third of the proton spin is due to quark
spins \cite{ellis,altar}. It is interesting that calculations of
$\eta '$ couplings to vector mesons  also predicted that quarks carry
about one third of spin of vector mesons \cite{band}.  These results
could be interpreted in such a way that a substantial part of hadron spin
would be due to an orbital angular momentum of quark matter.

It is also evident from deep--inelastic scattering data that
strange quarks play essential role in the structure of proton and in
its spin in particular. They are negatively polarized in polarized
nucleon, $\Delta s\simeq -0.1$.
Polarization effects in  hyperon production observed for the first time
  two decades ago also continue to
 demonstrate \cite{newhyp} that strange
 quarks produced in hadron interactions appear to be polarized.

In the recent paper \cite{asy} we considered a possible origin of asymmetry
in the pion production under collision of a polarized proton beam with
unpolarized proton target.  We have used a scheme which
incorporates perturbative and
nonpertur\-ba\-ti\-ve phases of QCD
and argued that the orbital angular momentum of partons inside
constituent quarks (retained by the perturbative phase of QCD under
transition from nonperturbative phase) leads to significant
asymmetries in hadron production with polarized beam.

 In this note we
 give an estimate for asymmetry in
 the  production of
 $\varphi$-meson.
 It is argued that this asymmetry is connected with orbital momenta of
 strange quarks entering the structure of constituent quarks.

 It is well known fact that mechanism of chiral symmetry
 spontaneous breaking
in QCD results, in particular, in generation of quark masses and in
appearance of quark condensates.  Quark masses are comparable
with the hadron
mass scale.  This supports a representation of hadron as a loosely bounded
system of the constituent quarks.  These observations on the hadron structure
lead to the understanding of static properties of hadrons and  several
regularities observed in hadron interactions at large distances.  However,
the structure of hadron depends on the scale of the process and is different
for different values of the scale $Q^2$.  Processes with large $Q^2$ would
resolve partonic structure of constituent quarks and can be described by
perturbative QCD.

 We consider a nonperturbative hadron to be consisting of the
  constituent quarks located at the central part of the hadron
and quark condensate surrounding this core.  Experimental and
theoretical arguments in favor of such picture were given,  e.g. in
\cite{isl,trtu}.  We refer to effective lagrangian approach and use
as a basis  the NJL model \cite{njl}.  The effective lagrangian  in
addition to the $4$--fermion interaction of the original NJL model
includes $6$--fermion $U(1)_A$--breaking term.  The constituent quark
masses have been calculated in \cite{jaffe}:  \begin{equation} m_U  =
m_u-2G\langle 0|\bar u u|0\rangle-2K\langle 0|\bar d d|0\rangle
\langle 0|\bar s s|0\rangle\label{ms} \end{equation} In this approach
massive  quarks appear  as quasiparticles, i.e. as current quarks and
the surrounding  clouds  of quark--antiquark pairs which consist of a
mixture of quarks of the different flavors.  Quark  radii are
  determined by the radii of  the  surrounding clouds.  Quantum
numbers of the constituent quarks are the same as the quantum numbers
of current quarks due to conservation of the corresponding currents
in QCD.  The only exception is the flavor--singlet, axial--vector
current, its $Q^2$--dependence is due to axial anomaly which arises
under quantization.  It is worth to stress that in addition to $u$
 and $d$ quarks constituent quark ($U$, for example) contains pairs
of strange quarks (cf. Eq. (\ref{ms})), and the ratio of scalar
density matrix elements \begin{equation} y=
{\langle U| \bar ss|U\rangle}
/ {\langle U|\bar u u+\bar d d+\bar s s|U\rangle}
 \label{str} \end{equation} can be estimated from different
approaches \cite{stein1} as $y=0.1 - 0.5$.
These estimates are based on the Feynman-Helmann theorem
$
{\langle Q_i| \bar q_jq_j|Q_i\rangle}=\partial M_i/\partial m_j.
$
Loosely speaking, $y$ can
  be considered as a probability to find $s$ and $\bar s$ quarks in
 the constituent quark \cite{ans}.

The picture of hadron consisting from
constituent quarks can be applicable at moderate momentum transfers,
while interactions at high momentum transfers would resolve internal
structure of constituent quarks and they are to be represented as
clusters of noninteracting partons in this kinematical region.  In
the framework of the NJL model the partonic structure of constituent
 quarks was defined in \cite{jaffe}.  Transition to partonic picture
 in this model is described by the introduction of a momentum cutoff
 $\Lambda=\Lambda_\chi\simeq 1$ GeV.  We adopt the point  that the
 need for such cutoff is an effective implementation of the short
 distance behavior in QCD.

   Spin of constituent quark $J_{U}$  in this approach is determined
 by the  following sum \begin{equation}
 J_{U}=1/2=J_{u_v}+J_{\{\bar q q\}}+\langle L_{\{\bar qq\}}\rangle.
 \label{bal} \end{equation} The value of the orbital
 momentum contribution into the spin of constituent quark can be
 estimated with account for new experimental results from
 deep--inelastic scattering \cite{altar} indicating that quarks carry
 one third of proton spin, i.e.  \[ (\Delta u +\Delta d +\Delta
 s)_p\simeq 1/3, \] and taking into account  the relation between
 contributions of current quarks into a proton spin and corresponding
contributions of current quarks into a spin of constituent quarks
and that of constituent quarks into  proton spin
\begin{equation} (\Delta u +\Delta d +\Delta s)_p = (\Delta U+\Delta
D) (\Delta u +\Delta d +\Delta s)_U.\label{qsp} \end{equation}
Indeed, if we adopt $SU(6)$ model ($\Delta U+\Delta D=1$) then we
should conclude  that $J_{u_v}+J_{\{\bar q q\}}\simeq 1/6$ and
   from Eq. (\ref{bal}) $\langle L_{\{\bar q q\}}\rangle\simeq
  1/3$, i. e. about 2/3 of the $U$-quark spin is due to the orbital
angular momenta of $u$, $d$ and $s$ quarks inside $U$-quark.

  The important point  what the origin of this orbital angular
  momentum is. It was proposed \cite{asy} to consider an analogy with
 an  anisotropic generalization of the theory of superconductivity
 which seems to match well with the above picture for a constituent
  quark.  The studies \cite{anders} of that theory show that the
   presence of anisotropy leads to axial symmetry of pairing
  correlations around the anisotropy direction $\hat{\vec{l}}$ and to
 the particle currents induced by the pairing correlations.  In
 another words it means that a particle at the origin is surrounded
 by a cloud of correlated particles which rotate around it with the
axis of rotation $\hat {\vec l}$.   Calculation of the orbital
momentum  shows that it is proportional to the density of the
correlated particles. Thus, it is clear that there is an analogy
between  this picture and that describing the constituent quark. An
axis of anisotropy $\hat {\vec l}$ can be associated with the
polarization vector of  valence quark located at the origin of the
constituent quark.  Then the orbital angular momentum $\vec L$ lies
along $\hat {\vec l}$ and its value is proportional to quark density
$ \langle Q|\bar q q|Q\rangle$.  A nonzero internal orbital
momentum of partons in the constituent quark means that there are
significant multiparton correlations.

 We argue that the existence of this orbital angular momentum, i.e.
 orbital motion of quark matter inside constituent quark, is the
 origin of the observed asymmetries in inclusive production at
  moderate and high transverse momenta.  Indeed, since the
 constituent quark has a small size, asymmetry associated with internal
 structure of this quark will be significant at
 $p_{\perp}>\Lambda_\chi\simeq 1$ GeV/c where interactions at short
 distances  give noticeable contribution.

 It should be noted that at high $p_\perp$ we will have a parton
 picture for constituent quark being a cluster of current quarks
 which however should naturally preserve  their orbital
 momenta of the preceding nonperturbative phase of QCD, i.e. the
 orbital angular momentum will be retained in perturbative phase of
 QCD and partons are to be correlated.

 Model for hadron production based on the above picture was described
 in \cite{asy}. We mention here its main points and consider hadron
  processes of the  type $ h_1^\uparrow +h_2\rightarrow \varphi +X
 $ with polarized beam or target. Interest in this process is
 related to the recently observed  OZI-rule violations
  (cf. \cite{rell} and referencies therein).
 According to the
 OZI rule such process containing pair of strange quarks in the final
 state should be suppressed compared to $\omega$-meson production,
 for example.

   The picture of hadron consisting of constituent quarks embedded
 into quark condensate implies that overlapping and interaction of
 peripheral clouds   occur at the first stage of hadron interaction.
 As a result massive virtual quarks appear in the overlapping region
 and  some effective field is generated.  Constituent quarks  located
 in the central part of hadron are supposed to scatter in a
 quasi-independent way by the effective field.  Inclusive production
  of $\varphi$-meson results  from the excitation of this constituent
 quark and  subsequent fragmentation of $\bar s s$-pair from this quark
 into $\varphi$ meson. This  process is determined by the distances
smaller than constituent quark radius and is associated therefore
 with hard interactions (high $p_{\perp}$'s).

 Explicit formulas for the corresponding inclusive cross--sections were
given in \cite{asy}. They have the form \begin{equation}
\frac{d\sigma^{\uparrow,\downarrow}}{d\xi}= 8\pi\int_0^\infty
bdb\frac{I^{\uparrow,\downarrow}(s,b,\xi)} {|1-iU(s,b)|^2},\label{un}
\end{equation} where $b$ is the impact parameter of colliding
particles and $\xi=(x,p_\perp)$. Here  function $U(s,b)$ is the generalized
reaction matrix (helicity nonflip one) which is determined by  dynamics of
the elastic reaction $h_1+h_2\rightarrow h_1+h_2$.  The functions
$I^{\uparrow,\downarrow}(s,b,\xi)$ are related to the functions $U_n
(s,b,\xi,\{\xi _{n-1}\})$ which are the multiparticle analogs of the
$U(s,b)$.  Arrows $\uparrow$ and $\downarrow$ denote corresponding
 direction of  transverse spin of the polarized initial  particle.

 We introduce the two functions $I_+$ and $I_-$:  \begin{equation}
I_{\pm}(s,b,\xi)=I^\uparrow(s,b,\xi)\pm I^\downarrow(s,b,\xi),
\end{equation} where the  function $I_+(s,b,\xi)$ is a
spin-independent one.  The following sum rule takes place for this
function:  \begin{equation} \int I_+(s,b,\xi)d\xi=\bar
n(s,b)\mbox{Im} U(s,b),\label{sr} \end{equation} where $\bar n(s,b)$
is the mean multiplicity of secondary particles in the impact
parameter representation.

Asymmetry $A_N$
can be expressed in terms of the functions $I_{\pm}$ and $U$:
\begin{equation} A_N=\frac{\int_0^\infty bdb
I_-(s,b,\xi)/|1-iU(s,b)|^2} {\int_0^\infty bdb
I_+(s,b,\xi)/|1-iU(s,b)|^2}.\label{xnn} \end{equation}

Taking into account Eq. (\ref{sr}), quasi-independence of the
 constituent quarks and assumption on hadron production as a result
of interaction of one of the  constituent quarks with the effective
field we adopt the following expressions for the functions
$I_+(s,b,\xi)$ and $I_-(s,b,\xi)$:  \begin{equation} I_\pm
(s,b,\xi)=\bar
n(s,b)\frac{1}{n_{h_1}}\mbox{Im}[\sum_{j=1}^{n_{h_1}}\prod_{i\neq
j}^{N} \langle f_{Q_i}(s,b)\rangle\langle
\varphi^\pm_{\varphi/Q_j}(s,b,\xi)\rangle], \label{prod}
\end{equation} where quarks $ Q_j$ are the constituent quarks in from
the hadron $h_1$.
 Factors $\langle f_{Q}(s,b)\rangle$ correspond to
the individual quark scattering amplitude smeared over transverse
position of $Q$ inside hadron $h$ and over fraction of longitudinal
 momentum of the initial hadron carried by quark $Q$.
 The functions
$\langle\varphi^\pm_{\varphi/Q_j}(s,b,\xi)\rangle$ describe the
  $\varphi$-meson production as a result of interaction of the
 constituent quark $Q_j$ with the effective field.
$N=n_{h_1}+n_{h_2}$ is the total number of constituent quarks in the
initial hadrons.

 The central point of the model is a connection  of the one-spin
 asymmetries in inclusive production with the orbital angular momentum
 of current quarks inside the constituent quark or in other words,
 with their coherent rotation.

 This orbital momentum
will affect the hadron production  at short distances where internal
structure of constituent quark could be probed. The functions
$\langle\varphi^\pm_{\varphi/Q}(s,b,\xi)\rangle$ will be sensitive to
internal structure of constituent quarks i.e. they will be determined
by the hard processes. These functions should account the presence
 of strange quarks inside the constituent quark and coherent
 rotation of current quarks and
 can be written as a convolution integral:
\begin{equation} \langle\varphi^\pm_{\varphi/Q}\rangle=
\langle\varphi^\pm_{{\bar s s} /Q}\rangle\otimes D_{\varphi/{\bar s
s}}, \end{equation} where $D_{\varphi/{\bar s s}}$ is the
fragmentation function which is supposed to be spin-independent.  The
 functions $\langle \varphi^\pm_{\bar s s/{ Q}}\rangle$ are determined
by the internal structure of constituent quark and include the
 structure function of constituent quark $\omega_{ s/Q}(x)$.

 The spin-dependent function
 $\langle \varphi^-_{\bar s s/{ Q}}\rangle$
 should account  the effect of nonzero orbital momentum of strange
 quark-antiquark pair
 (and consequently its internal transverse momenta $k_{\perp}$
 inside the constituent quark).  This internal orbital momenta
 means a certain shift in transverse momenta of produced pair,
 i.e. $p_\perp\rightarrow p_\perp+k_\perp$ and we suppose that
 the
  spin-dependent function
 $\langle \varphi^-_{\varphi/{ Q}}\rangle$ can be obtained
  by  shifting the transverse momenta
 in the spin-independent one
 $\langle \varphi^+_{\varphi/{ Q}}\rangle$.
 Since we consider rather high
 transverse momenta $p_\perp>1$ GeV/c,
 the effect of this shift will be reduced to phase factor in impact
 parameter representation, i.e. we will have a following relation
 \[
 \langle \varphi^-_{\varphi/{ Q}}(s,b,x,b')\rangle
 \simeq
\exp{[ i k_{\perp\bar s s/ Q} b']}
 \langle \varphi^+_{\varphi/{ Q}}(s,b,x,b')\rangle,
\]
where
\[
 \langle \varphi^+_{\varphi/{ Q}}(s,b,x,b')\rangle=
\int
d^2p_\perp
\exp{(i\vec{b}'\vec{p}_\perp)}
 \langle \varphi^+_{\varphi/{ Q}}(s,b,x,p_\perp)\rangle.
\]
 Then taking into account that quark matter distribution inside constituent
 quark has radius $r_Q$
  we can rewrite the phase factor in the form
\[
\exp{[ i k_{\perp\bar s s/ Q} r_Q]}=
\exp{[ i L_{\perp\bar s s/ Q}]}
\]
 and use the following approximate
relation between the functions $\langle\varphi^-_{\varphi/
Q}(s,b,\xi)\rangle$ and $\langle\varphi^+_{\varphi/ Q}(s,b,\xi)\rangle$:
\begin{equation}
\langle\varphi^-_{\varphi/ Q}(s,b,\xi)\rangle
 \simeq
\exp{[ i L_{\bar s s/Q}]}
\langle\varphi^-_{\varphi/ Q}(s,b,\xi)\rangle.
\label{*}
\end{equation}
 Since the  orbital angular momentum of $\bar s s$ quarks
in the constituent quark $ Q$ is proportional to its polarization and
 mean orbital momentum of strange quarks in the constituent quark, we
can rewrite the phase factor in Eq.(\ref{*}) in the form
$\exp{[i{\cal{P}}_{Q}\langle L_{\{s\bar s\}}\rangle]}$.  The mean
orbital momenta of strange quarks can be written in its turn as
follows \[ \langle L_{\{s\bar s\}}\rangle\simeq y\langle L_{\{q\bar
q\}}\rangle, \] where  the quantity $y$ is determined by Eq.
(\ref{str}). With account for above relations we can get from
 Eq. (\ref{xnn}) estimate for asymmetry $A_N$ in $\varphi$-meson
production at $p_\perp>1$ GeV/c:  \begin{equation}
A_N(\varphi)\propto \langle{\cal{P}}_{ Q}\rangle \langle L_{\{\bar q
q\}}\rangle y\simeq 0.01 - 0.05.  \label{an} \end{equation} For
polarization of the constituent quarks we  use $SU(6)$ values
${\cal{P}}_U=2/3$ and ${\cal{P}}_D=-1/3$. Value of orbital angular
momentum was taken to be $\langle L_{\{\bar q q\}}\rangle=1/3$ and
$y=0.1 - 0.5$ according to above estimates.

Thus, the model
predicts noticeable positive one-spin asymmetry at high $p_{\perp}$
values in inclusive $\varphi$-meson production.
This mechanism gives same magnitude of asymmetry $A_N(\varphi)$
 in $\bar p p$ interactions.
The above results show that it
 is reasonable to make experimental measurements of cross-section and
asymmetry in inclusive $\varphi$-meson production to study strange content of
constituent quark as a possible source of OZI-rule evasion.
Such measurements would provide information on the
 fraction of strange quarks inside constituent quark and their orbital
momenta.   \end{document}